\documentclass[traditabstract]{aa}
\usepackage{verbatim}
\usepackage{graphicx}

\newcommand{\sub}[1]{_{\rm#1}}
\usepackage{natbib,twoopt}
\newcommand{\npix}{{n\sub{pix}}}
\newcommand{\nmodpix}{{n\sub{modpix}}}
\newcommand{\nim}{{n\sub{im}}}
\usepackage{txfonts}
\bibpunct{(}{)}{;}{a}{}{,}

\begin{document}

   \title{Direct model fitting to combine dithered ACS images}

   \author{H. Mahmoudian
          \inst{1}\thanks{\email{hanieh@astro.uni-bonn.de}}
          \and
          O. Wucknitz\inst{1,2}
          }

   \institute{Argelander-Institut f\"ur Astronomie, Universit\"at Bonn \\
  			  Auf dem H\"ugel 71, 53121, Bonn, Germany.
         \and
             Max-Planck-Institut f\"ur Radioastronomie,\\
             Auf dem H\"ugel 69, 53121, Bonn, Germany.
             }

   \date{Received  22 February 2013; accepted 10 July 2013}

  \abstract{The information lost in images of undersampled CCD cameras
    can be recovered with the technique of `dithering'. A number of
    subexposures is taken with sub-pixel shifts in order to record
    structures on scales smaller than a pixel. The standard method to
    combine such exposures, `Drizzle', averages after reversing the
    displacements, including rotations and distortions. More
      sophisticated methods are available to produce, e.g., Nyquist
      sampled representations of band-limited inputs. While the
      combined images produced by these methods can be of high
      quality, their use as input for forward-modelling techniques in
      gravitational lensing is still not optimal, because the residual
      artefacts still affect the modelling results in unpredictable
      ways.  In this paper we argue for an overall modelling approach
      that takes into account the dithering and the lensing without
      the intermediate product of a combined image. As one building
      block we introduce an alternative approach to combine dithered
    images by direct model fitting with a least-squares approach
    including a regularization constraint. We present tests with
    simulated and real data that show the quality of the
    results. The additional effects of gravitational lensing
      and the convolution with an instrumental point spread function
      can be included in a natural way, avoiding the possible
      systematic errors of previous procedures.}


   \maketitle

\section{Introduction}

The light distribution detected by a CCD in an astronomical
observation does not resemble the brightness distribution on the sky
exactly but is affected by the properties of the optical
instrument. The most important effect is the point-spread-function
(PSF). If we neglect small variations of PSF with position, it can be
described as a convolution of the true light distribution. The effect
is caused by diffraction in the instrument and, for ground-based
telescopes, by the atmospheric `seeing'.

In addition, the design of the instrument can cause a geometric
distortion.  In the case of the Advanced Camera for Surveys (ACS) of
the Hubble Space Telescope (HST), the off-axis location of the
detectors (tilted focal surface with respect to the light rays) and
the Optical Telescope Assembly (OTA) cause anisotropic variations in
the plate scale. The first ACS geometric distortion campaign observed
the core of 47 Tucanae with the WFC (Wide Field Channel) and HRC (High
Resolution Channel). A fit of a 4th order polynomial corrects the
distortion to an accuracy of 0.1--0.2 pixels over the entire field of view 
\citep{acshb}.

Each pixel integrates the light over its area (and slightly beyond)
weighted with the `pixel response function', assumed to be the same
for all pixels. This can mathematically be described as a convolution
with this response function and subsequent multiplication with a
`bed-of-nails' sampling function.

Therefore if we consider the true light distribution as $I^*$, we can
write the observed light distribution as
\begin{equation}
I^{\textrm{obs}}=S\cdot R\star D \otimes P \star I^*.
\label{observation}
\end{equation}
Here $P$ represents the PSF, $D$ the effect of optical distortion, $R$
the response function and $S$ the sampling function. A convolution is
denoted by $\star$, an arbitrary mapping by $\otimes$ and a
multiplication by a dot. For our image analysis we either have to
`invert' this process, by trying to construct a realistic representation, e.g.\ band-limited, of the input image, or take it into account in
a modelling process.

In order to reveal information on scales smaller than the pixel size,
observations are `dithered', i.e. split into several exposures that
are shifted relative to each other by sub-pixel
displacements. Inverting the convolution with $P$ requires the
conservation of structures on scales smaller than the PSF width; the
sampling must be sufficiently fine to resolve the PSF. For
ground-based telescopes, $P$ is dominated by the atmospheric seeing,
and modern CCD instruments generally have sufficient resolution. For
the WFC of ACS, the pixel size of 0.05 arcseconds is comparable to the
overall PSF width, and dithering is essential to recover any of the
fine-scale structure of the PSF.

\subsection{Image combination and lens modelling}

There are a number of known methods that can invert (in the
  sense described above) the measurement process and produce an
  estimate of the true light distribution under certain assumptions
  for its properties (see below). Our main interest are gravitational
  lenses, and here the main goal is not a good estimate of the true
  (gravitationally lensed) sky brightness distribution. Instead we
  want to derive good models for the \emph{unlensed} brightness
  distribution of the source and in particular good models for the
  mass distribution that causes the lensing effect.

Sophisticated methods are available that can invert the effect of the
lens mapping in order to reconstruct the unlensed source and to
determine the mass distribution of the lens
\citep[e.g.][]{invert2,invert3,invert1}. However, they all rely on an
accurate input image that is free of systematic artefacts. Because of
the complexity of the overall problem, the effect of artefacts in the
input image on the final result are not easy to judge.  Even the best
reconstruction methods will necessarily introduce errors into the
reconstructed $I^*$, and the optimal way of doing the lens modelling
is not to first make an optimal image and then use that as basis for
the mass models of the lens. Instead we eventually want to combine
both parts and include the lens effect as part of the measurement
process and fit for the true brightness distribution of the source and
the mass model of the lens, without the intermediate result of
  a combined image. Even though the final reconstructed image of the
unlensed source will still suffer from minor systematic effects, the
effect on the mass models will be less severe and much better
understood.

In the notation of Eq.~\eqref{observation} the lens acts as another
generally complicated mapping of the unlensed brightness distribution
to form $I^*$. This mapping is linear in the brightness distribution
but depends non-linearly on the lens model parameters. The final
algorithm will thus consist of two nested parts, where the inner one
determines the best brightness distribution of the source for a given
lens model, while the outer one will vary the lens model itself to
minimize the residuals.

This paper describes the first step in this direction,
  consisting of an algorithm that solves parts of
  Eq.~\eqref{observation} by explicitly modelling the optimal true
  brightness distribution that fits the data, taking into account the
  dithering with full correction for distortion, and at the same time
  obeys the necessary regularization constraints.

Some of the image combination methods described below
   reconstruct a Nyquist sampled representation of the observed sky
   brightness distribution that has to be band-limited as result of
   the finite instrumental aperture. Such an approach is not
   appropriate in our case, because the lens mapping will destroy this
   property for the brightness distribution of the source. Instead we
   employ a smoothness regularization constraint.

We emphasize that out modelling algorithm is not meant to
  replace other methods to produce combined images, but to provide a
  major part of a full algorithm that will include the action of the
  lens and the convolution with the PSF. As test of this part we
  nevertheless use the combined images as a demonstration.

\subsection{Known image combination methods}

A basic method to combine these dithered images is a linear technique,
`shift-and-add'. In this method, pixels of each image are transferred
to a finer grid, shifted to the same position and added to the output
image. In the formalism introduced above this corresponds to a
convolution with another function $R'$ that represents the size and
shape of the pixels. Neglecting the distortion for the moment, and
assuming a uniform and complete dithering pattern (so that the
sampling $S$ can be neglected), the result would correspond to
$R'\star R \star P \star I^*$. This additional convolution with the
pixel size reduces the resolution of the result. The geometric
distortion is generally not taken into account in this simple process.

`Drizzle', the standard method to combine dithered HST images, extends
this approach by allowing for arbitrary shifts and rotations, by
correcting for the distortion and by using more sophisticated
techniques to transfer the images to finer grids. The general
approach, however, is based on shift-and-add, and Drizzle always
introduces a convolution function $R'$ \citep{drizzle}.

Drizzle adds small high-frequency artefacts to the image (Figure 1 in
\cite{idrizzle} shows this effect) which is detrimental in cases where
preservation of the true PSF is needed. On larger scales (larger than
the original pixels) these artefacts are averaged out. More details
are provided in \cite{idrizzle}.

With nearly interlaced
  dithered images, high-fidelity combined images can be created with the
  method introduced by \cite{lauer}.  This method attempts to predict
  the values of the output image based on dithered data but it can not
  handle image distortions. It is based in the Fourier transform of
  the images. The final image is computed as a linear combination of
  the transformed input images. In this process the aliased components
  are removed algebraically.

\cite{rowe} present a method (Imcom)
  to combine undersampled images for band-limited data. Imcom
  is developed to process data based on linear algorithm
  approach in Fourier space. In this method, the assumed function for
  the PSF (including jitter, optics of the instrument and the response of
  the CCD) is specified for each pixel of the images. The required
  information to derive the function comes from the optical model for
  the instrument and/or stars in the field. The fully parametric model
  can be adopted for this PSF function.

\cite{idrizzle} introduce the (iDrizzle) method, which is
  based on Drizzle but removes high frequency artefacts using the
  assumption of a band-limited image. The iterative Drizzle approach
  also eliminates the convolution with an additional kernel that is
  inherent in Drizzle.

The available methods have in common that they are designed to produce
a combined image without direct forward modelling. They are thus not
easily useable as part of the full lens modelling process.

\section{The direct fitting method}

The development of an alternative combining method was prompted by the
case of the gravitational lens system \object{B0218+357} \citep{patnaik}. With
334 mas it has the smallest image separation of all known galaxy-scale
lenses, and very accurate positions of the images relative to the
lensing galaxy are required for a reliable determination of the Hubble
constant from the time delay \citep{biggs1999b0218+}.  \cite{york}
used Drizzle to combine the data to determine the position of the galaxy relative to the
lensed images from deep ACS observations. They were able to improve
the determination of $H_0$ in this way, but the accuracy of the result
is limited by imaging artefacts and PSF inaccuracies.

Here we introduce an alternative method to combine the observed
images. This method is based on fitting a brightness distribution to
the observed exposures with a least-squares method.  Since we want to
overcome the undersampling problem, we use a finer grid for the
model image. A smoothness constraint is employed in order to obtain a unique
and realistic solution for the final result. Because the resolution in
the output image is higher than in the original input images, it is
obvious that such an additional constraint is required to avoid
unconstrained output pixels.

The model is pixelized in the sky coordinates, and in the
  transformation between CCD and sky coordinates the geometric
  distortion correction is taken into account. We use a simplified
(currently without explicit convolution with the PSF and
pixel-response function) version of the imaging equation
(\ref{observation}) to predict the observed brightness distributions
in CCD coordinates for a given model. The deviations of these
predictions from the real observations are then minimized by varying
the model pixel values.

The function that is being minimized can be written as
\begin{equation}\label{5}
f=\sum_{j=1}^\nim\sum_{i=1}^\npix\frac{\left(I^{\textrm{mod}}_{\textrm{int}(j)}[i]-I^{\textrm{obs}}_{(j)}[i]\right)^2}{\sigma_{ij}^2}w_{ij}+\lambda R(n)
\end{equation}
where $\nim$ is the total number of images that we want to combine,
$\npix$ is the number of CCD pixels per image, $I^{\textrm{mod}}$ is
the brightness distribution model in sky coordinates, and
$I_{(j)}^{\textrm{obs}}$ the observed image $j$. The subscript
`int$(j)$' denotes interpolation to the same grid as the observed
image $j$, taking into account the distortion, shift and rotation in
the conversion between CCD coordinates and sky coordinates.
$\sigma_{ij}$ is the uncertainty of pixel $i$ in image $j$.  The
additional weight function $w_{ij}$ is set to zero for flagged or
masked bad or unwanted pixels and to one otherwise. Most important is
cosmic ray flagging using the data quality layer of the calibrated and
flat fielded data (flt images).  The strength of the smoothness
constraint is given by the coefficient $\lambda$. More details on
choosing this parameter can be found in section \ref{lambda}.

$R(n)$ is a quadratic operator that measures non-smoothness, where $n$
denotes the order of derivatives included. Most commonly used are
gradient minimization ($n=1$) and curvature minimization ($n=2$):
\begin{eqnarray}\label{r1}
R(1)&=&\sum_{i=1}^\nmodpix \left[ \left( \frac{\partial
      I^{\textrm{mod}}}{\partial x}[i]\right)^2 + \left( \frac{\partial
      I^{\textrm{mod}}}{\partial y}[i]\right)^2 \right]
\\
\label{r2}
R(2)&=&\sum_{i=1}^\nmodpix\left( \frac{\partial^2 I^{\textrm{mod}}}{\partial x^2}[i]+\frac{\partial^2 I^{\textrm{mod}}}{\partial y^2}[i] \right)^2.
\end{eqnarray}
Here we are summing over all $\nmodpix$ model pixels which is
generally different from $\npix$. The derivatives of the discrete
model brightness distributions are determined using finite
differences.

In our work we use $n=1$, but in order to study the effect of
smoothing on our model, we also considered the case of $n=2$.
The function $f$ is quadratic in the unknown parameters so
  that the mininization corresponds to solving a system of linear
  equations. This could in principle be done directly. However, this
  is not trivial given the size of the system. For a $(1{\rm k})^2$
  model image, the matrix that has to be inverted has $10^{12}$
  elements. Instead we use the more general numerical minimization
  scheme named after Broyden, Fletcher, Goldfarb \& Shanno
  (BFGS). This quasi-Newtonian algorithm \citep{Press:1992:NRC:148286}
  uses analytical gradients and builds up information on the Hessian
  matrix iteratively. We use a limited-memory version (L-BFGS) that
  does not require storage of the full matrix and allows us to process
  the data on normal desktop computers.

In order to study how the smoothness constraint affects the combined
output image, we describe the effect as a convolution. If we assume
small pixels (large $\npix$ and $\nmodpix$) and $\sigma=1$, we can write
Eq. (\ref{5}) with continuous integrals as
\begin{equation}
f=\iint \left[I^{\textrm{mod}}(x,y) -
  I^{\textrm{obs}}(x,y)\right]^2\,\mathrm{d}x\,\mathrm{d}y+\lambda R(n)
\end{equation}
with
\begin{eqnarray}
R(1) &=&\iint \left[ \left( \frac{\partial I^{\textrm{mod}}}{\partial
      x}\right)^2 + \left( \frac{\partial I^{\textrm{mod}}}{\partial
      y}\right)^2 \right] \mathrm{d} x\, \mathrm{d} y  , \\
R(2) &=&\iint \left( \frac{\partial^2 I^{\textrm{mod}}}{\partial
      x^2} + \frac{\partial^2 I^{\textrm{mod}}}{\partial
      y^2}\right)^2 \mathrm{d} x\, \mathrm{d} y .
\end{eqnarray}
This approach is accurate in the limit of fine sampling of
$I^{\textrm{mod}}$ and dense dithering patterns. Considering
Parseval's theorem, we can transfer this equation to Fourier space
which makes the calculation simpler. This leads to a uniform
expression for \mbox{$n=1,2$},
\begin{eqnarray}\label{2}
  f=\iint \left|\widehat{I^{\textrm{mod}}}(u,v) -
    \widehat{I^{\textrm{obs}}}(u,v)\right|^2\,\mathrm{d}u\,\mathrm{d}v+
\nonumber \\
\lambda'\iint
  k^{2n}\left|\widehat{I^{\textrm{mod}}}(u,v)\right|^2\,\mathrm{d}u\,\mathrm{d}v ,
\end{eqnarray}
where $k^2=u^2+v^2$ and $\lambda'=(2\pi)^{2n}\lambda$.  Minimizing $f$
requires that the integrand is minimized, which results in
\begin{equation}
\widehat{I^{\textrm{mod}}}(u,v)=\frac{\widehat{I^{\textrm{obs}}}(u,v)}{1+\lambda' k^{2n}}.
\end{equation}
From Fourier theory we know that the multiplication of two functions
in Fourier space corresponds to the convolution of those functions in
image space. We can determine this convolution function $C(x,y)$ from
the inverse transform of $1/(1+\lambda k^{2n})$
\begin{equation}
C(x,y)=\iint\frac{1}{1+\lambda' k^{2n}}\mathrm{e}^{2i\pi\vec{k}\cdot\vec{x}} \,\mathrm{d}u\,\mathrm{d}v \, , \, \vec{k}=(u,v) .
\end{equation}

\begin{figure}[ht]
\centering
\resizebox{\hsize}{!}{\includegraphics{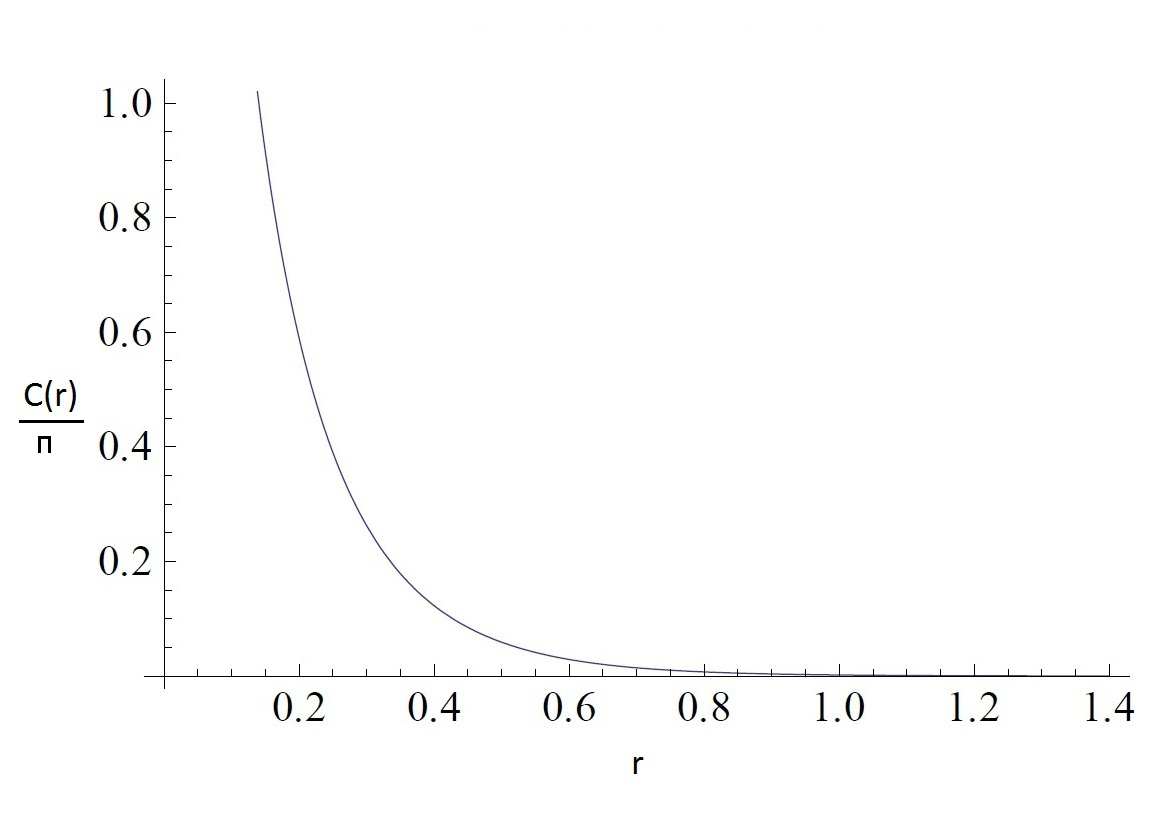}}
\caption{Convolution function $C(r)$ for gradient minimization ($n=1$)
  with $\lambda'=1$. The singularity at the origin is logarithmic.}
\label{K0}
\end{figure}

In the following we study this inverse transform for $n=1$ and $n=2$.
\paragraph{Gradient minimization, $n=1$:}
We invert the Fourier transform in spherical coordinate with
\mbox{$(x,y)=r(\cos\phi,\sin\phi)$} and \mbox{$(u,v)=k(\cos\theta,\sin\theta)$}.
\begin{eqnarray}
C(r)&=&\frac{1}{\lambda'}\int\limits_{0}^{\infty}\int\limits_{0}^{2\pi}
\frac{k}{\frac{1}{\lambda'}+k^2}\mathrm{e}^{2i\pi kr\cos(\theta-\phi)}
\,\mathrm{d}k\,\mathrm{d}\theta \\
&=&\frac{2\pi}{\lambda'}\int\limits_{0}^{\infty} \frac{k}{\frac{1}{\lambda'}+k^2}\mathrm{J}_{0}(2\pi kr) \,\mathrm{d}k
\end{eqnarray}
This integral is the definition of the \textit{Hankel} transform of
$(1/\lambda'+k^2)^{-1}$, which leads us to
\begin{equation}
C(r)=\frac{2\pi}{\lambda'}\mathrm{H}\left\lbrace \frac{1}{\frac{1}{\lambda'}+k^2}
\right\rbrace .
\end{equation}
The Hankel transform of this function is the modified Bessel function
of the second kind \citep{book}. Hence the convolution
function becomes
\begin{equation}
C(r)=\frac{2\pi}{\lambda'}\mathrm{K}_{0}\left( \frac{2\pi r}{\sqrt{\lambda'}} \right).
\end{equation}
The behaviour of this convolution function is shown in Figure \ref{K0}.

\begin{figure}[ht]
\centering
\resizebox{\hsize}{!}{\includegraphics{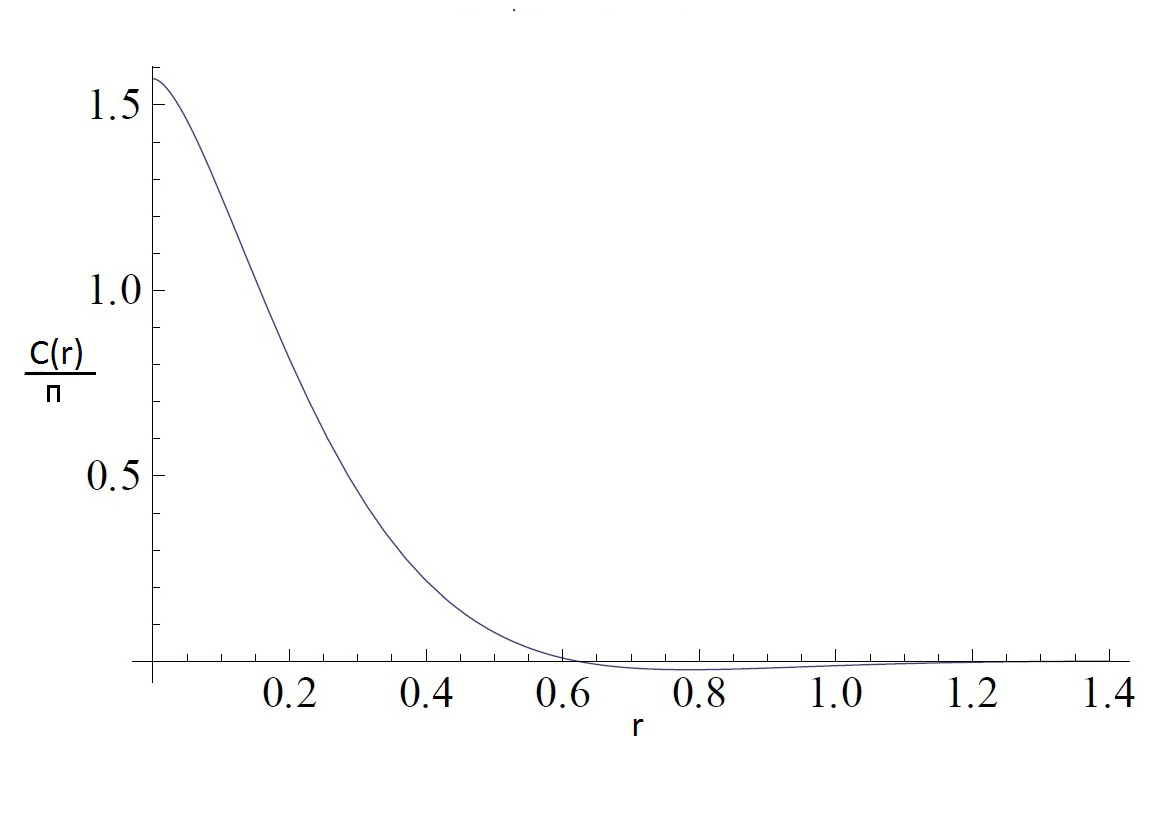}}
\caption{Convolution function $C(r)$ for curvature minimization
  ($n=2$) with $\lambda'=1$. In contrast to the case of $n=1$ this
  function is regular at the origin.}
\label{n2}
\end{figure}

\paragraph{Curvature minimization, $n=2$:}
For this case, the convolution function becomes 
\begin{equation}
C(r)=\frac{2\pi}{\lambda'}\mathrm{H}\left\lbrace\frac{1}{\frac{1}{\lambda'}+k^4}\right\rbrace .
\end{equation}
This results in the \textit{Kelvin} function $\mathrm{Kei}_{0}$
\citep{book}:
\begin{equation}
C(r)=\frac{-2\pi}{\sqrt{\lambda'}}\mathrm{Kei}_{0}\left( \frac{2\pi r}{\sqrt[4]{\lambda'}} \right)
\end{equation}
Figure \ref{n2} presents the behaviour of this convolution function.

\section{Tests of the direct fitting method}\label{lambda}

In our current implementation the algorithm consists of two
  separate parts. A Python script is used to calculate the mapping
  between CCD pixels and sky coordinates using the geometric
  distortions with parameters derived from the available
  meta-data. Particularly between different HST `visits', the
    dither offsets from the FITS headers are generally not
    sufficiently accurate. In this case they have to be determined
    using cross-correlation techniques or by using the positions of
    reference stars. In the results presented here, this was not necessary.
 Information about the mapping is added to the FITS images
  which are then read by the main minimization program written in
  C. For the minimization we use the GSL (GNU Science Library) variant
  of L-BFGS.\footnote{\url{http://www.gnu.org/software/gsl/}}

In this section we present the combined images of simulated and real
data based on our method. As real data we use observations of the two
gravitational lensing systems \object{B1608+656} and B0218+357.

\subsection*{Simulated Data}

\begin{figure}[ht]
\centering
\resizebox{\hsize}{!}{\includegraphics{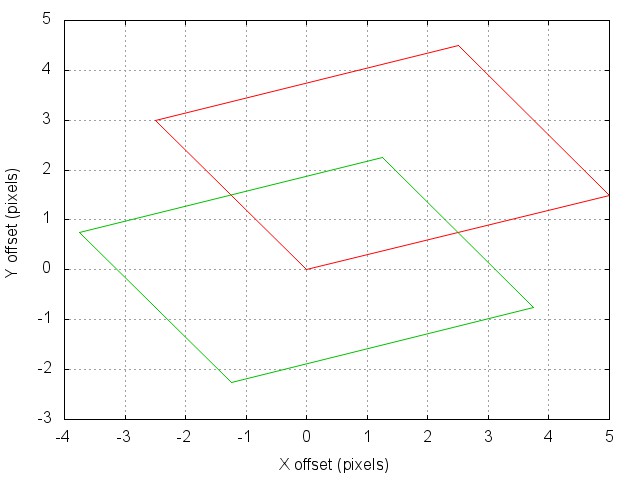}}
\caption{Box dither pattern used for the simulated data and real data.}
\label{dither}
\end{figure}

For the simulated data, we produced 20 images (a dithering pattern of
the ACS as shown in figure \ref{dither} was used in the simulation)
consisting of two sources with Gaussian brightness distributions at
fixed separation. This setup was chosen to roughly resemble the case
of B0218+357 in which relative positions are the most important
parameter.  Realistic Poisson noise was added to the simulated images.

\begin{figure}[ht]
\centering
\resizebox{\hsize}{!}{\includegraphics{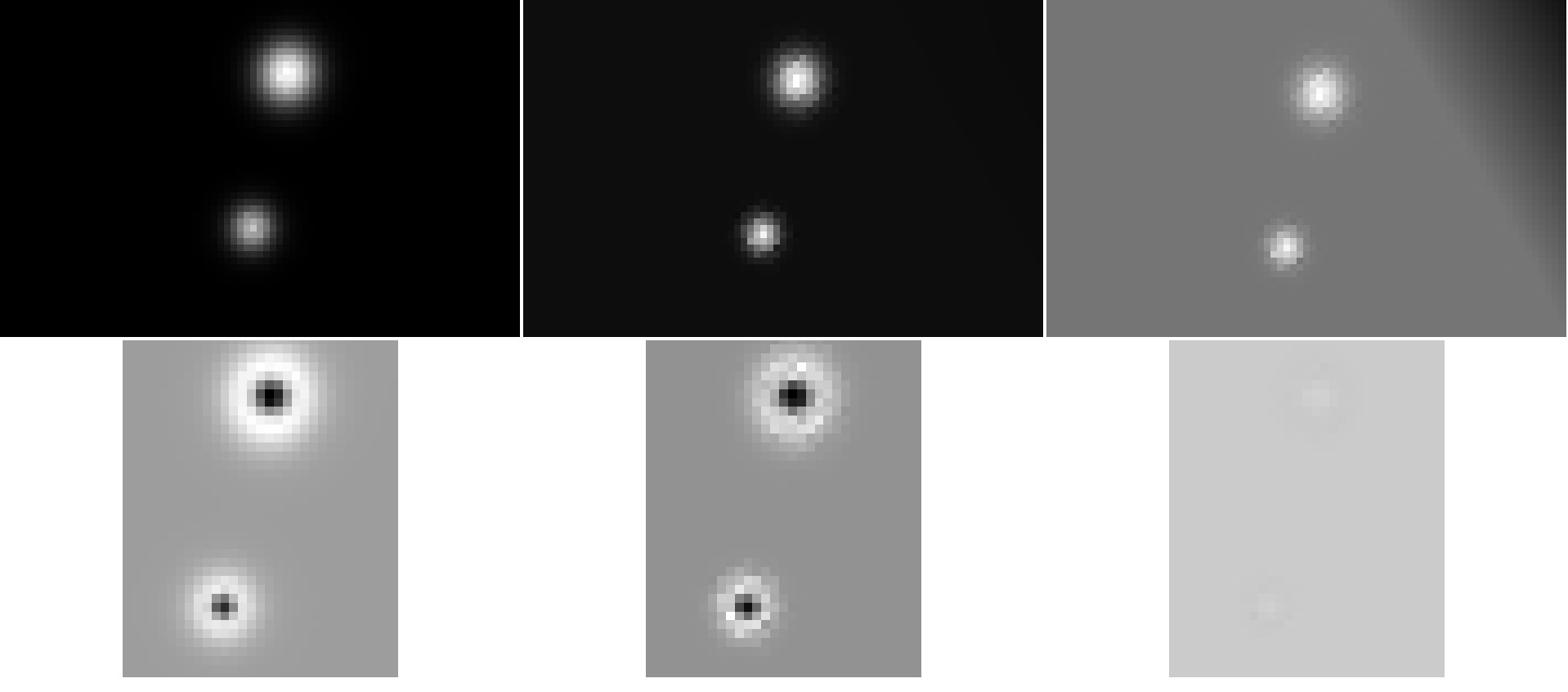}}
\caption{Top: Fitting results for the simulated images using smoothing
  coefficients of $\lambda=10^{-1}$, $10^{-3}$ and $10^{-5}$.  Bottom:
  Corresponding residuals relative to the input image.}
\label{sim2}
\end{figure}

The simulated images were combined with our method using a model pixel
size of 30 mas. We then used \textit{galfit} \citep{galfit} to fit the
source parameters as a test of the reconstruction accuracy.
To study the effects of a varying smoothing coefficient we used these simulated
data and combined them with different $\lambda$ (Fig.~\ref{sim2}).

Table \ref{tablei} presents the fitted separation of the sources in
the combined image for different values of $\lambda$.
\begin{table}[hbtp]
\centering
\caption{Comparison between the true separation of 0.72358 and the
  separation determined by galfit.}
\begin{tabular}{c c c}
\hline\hline
$\lambda$ & separation [arcsec] & error of separation [arcsec] \\
\hline
$10^{-1}$ & 0.7268 & 0.0032 \\
$10^{-2}$ & 0.7260 & 0.0025 \\
$10^{-3}$ & 0.7257 & 0.0022 \\
$10^{-4}$ & 0.7251 & 0.0015 \\
$10^{-5}$ & 0.7238 & 0.0002 \\
\hline
\end{tabular}
\label{tablei}
\end{table}

As can be seen from the table, $\lambda=10^{-5}$ provides the best
determination of the distance between the two simulated sources. For
real observations, the coefficient $\lambda$ can be determined by
comparing the $\chi^2$ of the fits for different values of $\lambda$
and selecting the one that is closest to the statistical expectation.

We conclude that the direct fitting approach can produce combined
images that in principle allow relative position measurements with an
accuracy well beyond 1mas.

\subsection*{B1608+656}
As a first test of the method on real data, we used the strong lensing
system B1608+656. We chose an observation consisting of 44 images
observed in the F814W filter with the ACS/WFC.

\begin{figure}[ht]
\centering
\resizebox{\hsize}{!}{\includegraphics{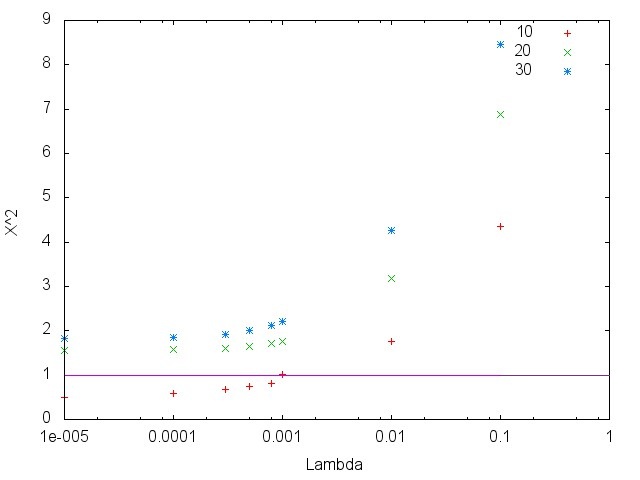}}
\caption{Reduced $\chi^2$ as a function of $\lambda$ for
  B1608+656. Different colours and symbols represent pixel sizes from
  10 to 30 mas.}
\label{lambda-psize-b1608}
\end{figure}

We applied the direct fitting method with different smoothing
coefficients and pixel sizes to find the optimal $\lambda$ for these
data. Figure \ref{lambda-psize-b1608} shows the variation of $\chi^2$
as a function of $\lambda$ for different pixel sizes.

For a pixel size of 10 mas we find that $\lambda=10^{-3}$ results in a
reduced $\chi^2$ of 1.029, very close to the expected 1. Generally
smaller values for $\lambda$ produce a better fit, as
expected. However, for pixel sizes of 20 and 30 mas, $\chi^2$ never
drops below unity. In these cases the large pixels themselves serve as
additional regularization that is too strong to achieve a reduced
$\chi^2$ of unity.

\begin{figure}[ht]
\centering
\resizebox{\hsize}{!}{\includegraphics{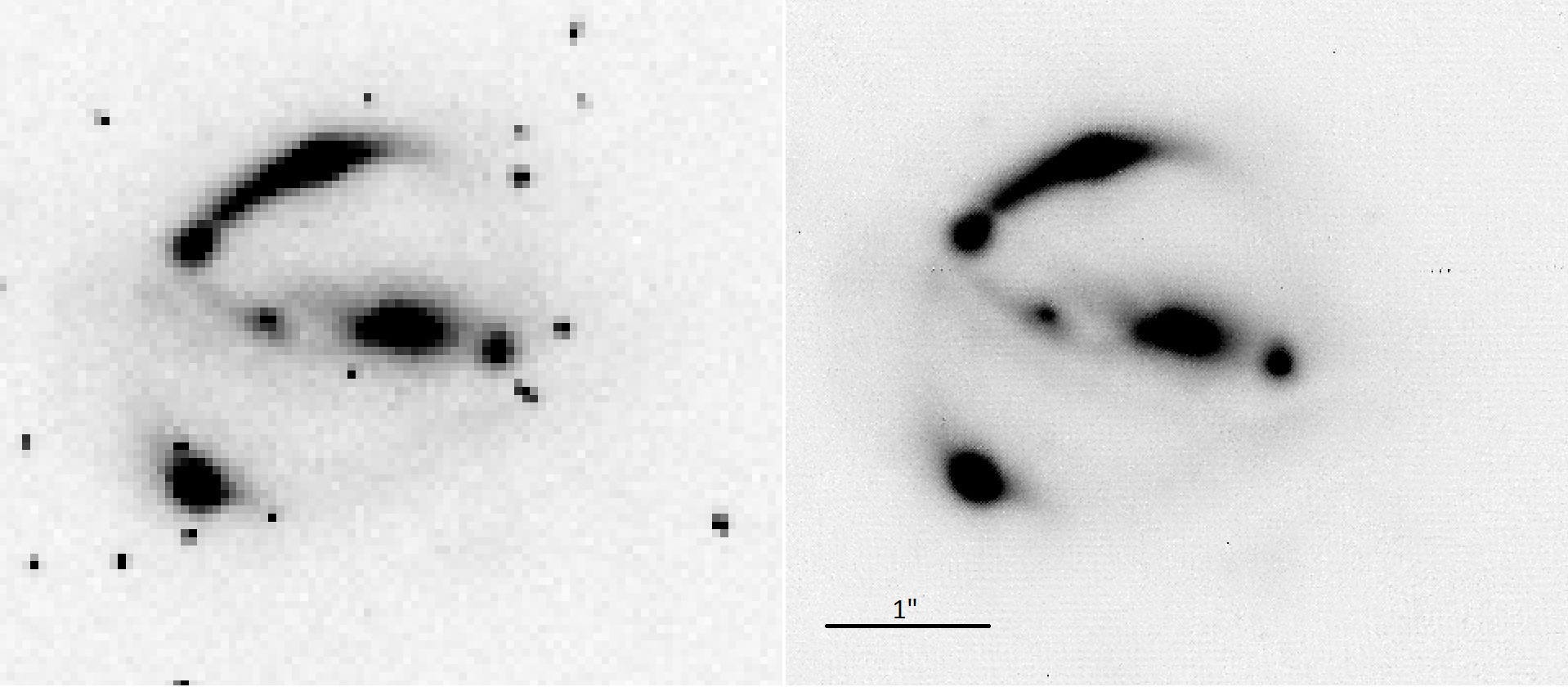}}\\
\resizebox{\hsize}{!}{\includegraphics{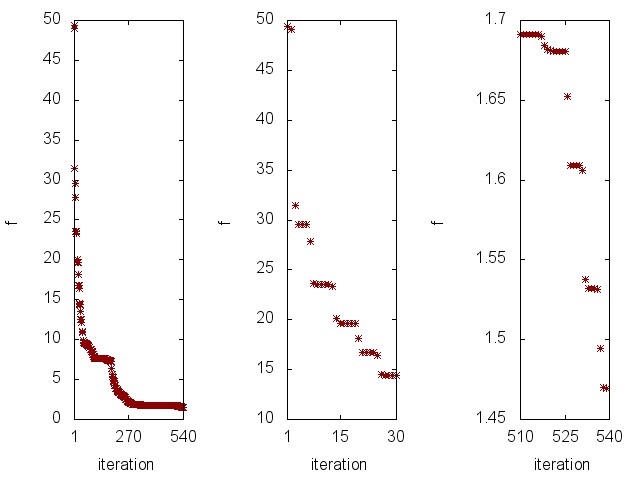}}
\caption{Top: One of the raw images of B1608+656 and combined image
  using the direct fitting method with $\lambda=10^{-3}$ and pixels of
  10 mas.  Bottom: Progress of the minimization of $f$ for all
  iterations, for the first iterations and for the final
  iterations. The final reduced $\chi^2$ for the combined image is
  $1.029$.}
\label{B1608}
\end{figure}

Figure \ref{B1608} shows one of the raw images and the combined image
using a pixel size of 10 mas and the optimal $\lambda=10^{-3}$.

\subsection*{B0218+357}
A second test was performed for the system B0218+357 that provided our
main motivation for the development of the fitting method.

\begin{figure}[ht]
\centering
\resizebox{\hsize}{!}{\includegraphics{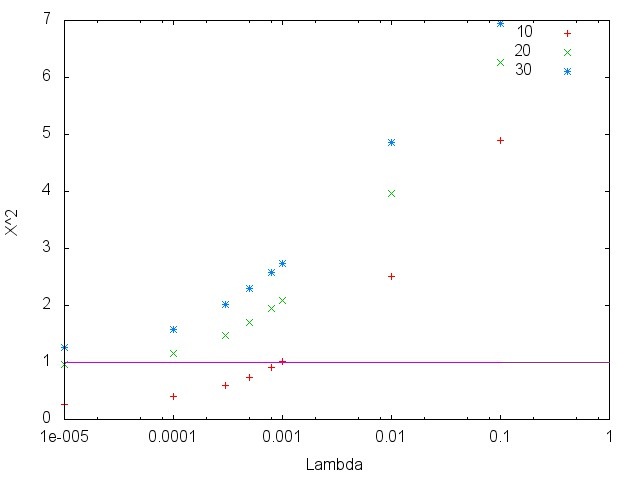}}
\caption{Reduced $\chi^2$ as function of $\lambda$ for
  B0218+357. Different colours and symbols represent pixel sizes from
  10 to 30 mas.}
\label{lambda-psize-b0218}
\end{figure}

\begin{figure}[ht]
\centering
\resizebox{\hsize}{!}{\includegraphics{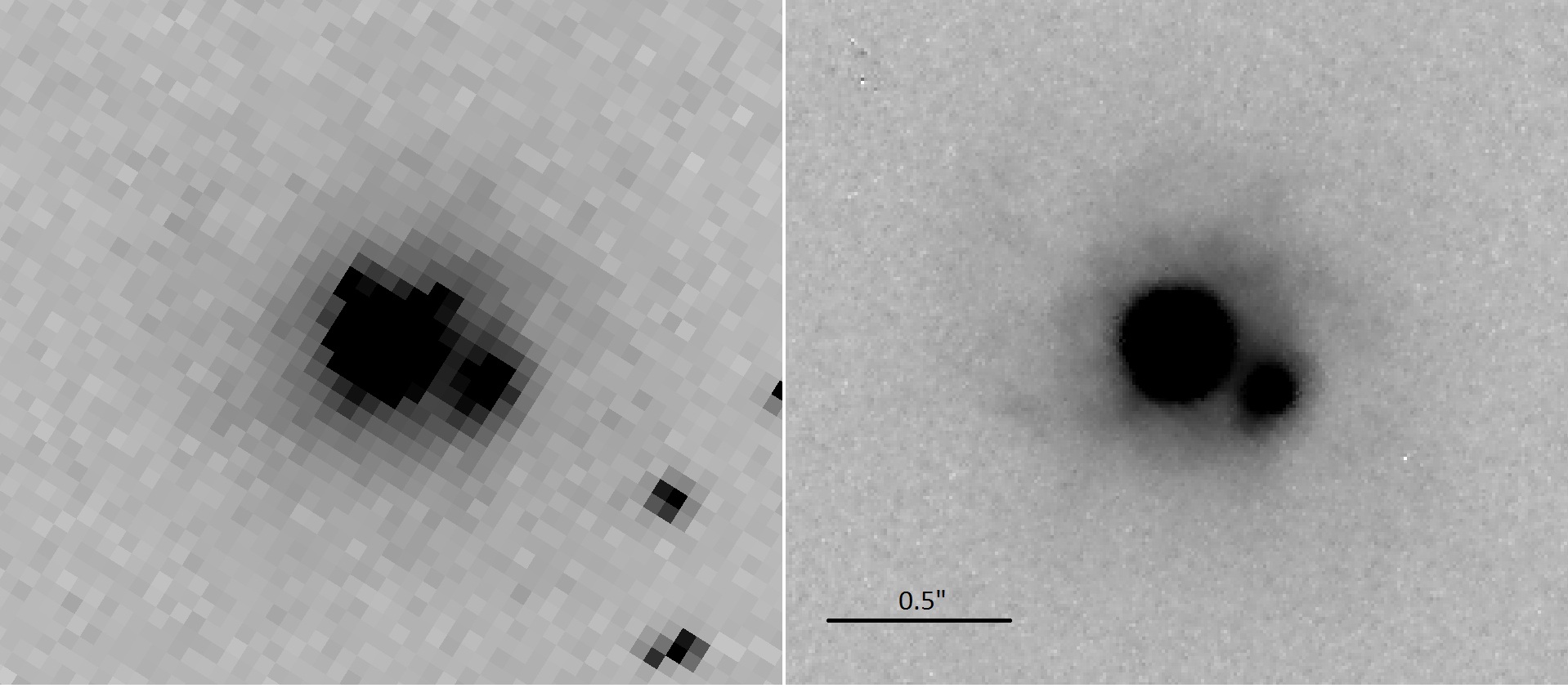}}\\
\resizebox{\hsize}{!}{\includegraphics{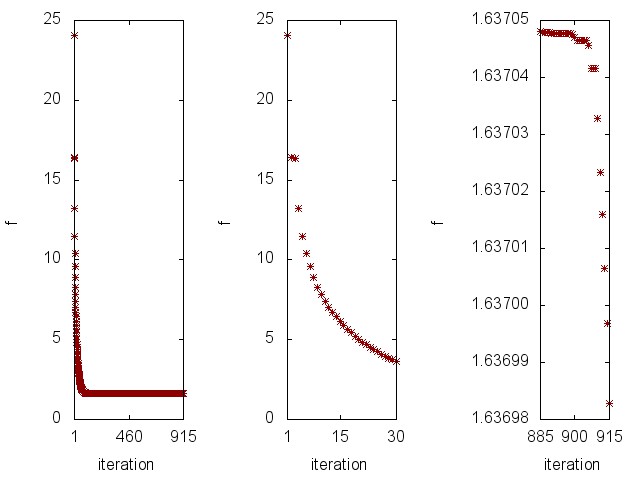}}
\caption{Top: One of the raw images of B0218+357 and combined image
  using a pixels size of 10 mas and $\lambda=10^{-3}$. Bottom:
  Progress of the minimization of $f$ for all iterations, for the
  first iterations and for the final iterations.}
\label{B0218}
\end{figure}

Of the observations described by \cite{york} we restricted ourselves
to visit 13, containing 20 images taken with ACS/WFC and the F814W
filter.
In order to determine the best $\lambda$ we used the same procedure as
for B1608+656 and plotted the resulting $\chi^2$ for a range of
$\lambda$ and pixel sizes (Fig.~\ref{lambda-psize-b0218}). We find
that $\lambda=10^{-3}$ with $\chi^2=1.008$ is close to the optimum. A
pixel size of 30 mas is again too large for a good fit, but 20 mas
would just be sufficient for a smaller $\lambda$.
Figure~\ref{B0218} shows the combined image resulting from the direct
fitting method with a pixel size of 10 mas.

\section{Conclusions and outlook}

In observations with undersampling cameras, many exposures with
sub-pixel shifts (dithering) are usually taken to record information
on scales smaller than a pixel. Drizzle and more sophisticated
  methods can be used to recover this information and produce combined
  images of high quality. However, even the best methods cannot avoid
  artefacts in the combined images, implying that such images are
  not well suited as input for further model fitting, e.g. in the
case of gravitational lenses where the true (unlensed) brightness
distribution of the source as well as the mass distribution causing
the lensing are to be determined.  In such a situation, a direct
model-fitting approach is much more appropriate to tackle this inverse
problem.

In this paper we presented a basic building block for such a fitting
procedure. Instead of trying to combine the dithered images by
shifting and adding, or by constructing a representation that
  is band-limited and consistent with all input images, we directly
fit the true brightness distribution of the sky to the observed
images. In this process we correct for the geometric distortion and
take possible flagging into account. A smoothness constraint is added
to allow for a unique solution and to avoid unrealistic small-scale
fluctuations and noise amplification. This approach of direct
  model fitting has the advantage that the additional mapping caused
  by a gravitational lens can later be included easily.

Tests with real and simulated data show that the fitting method works
well and produces accurate results. In the future we will use combined
images of B0218+357 to determine the relative positions of the lensed
images and the lensing galaxy with optimal accuracy. This is an
essential input for modelling efforts that are needed to convert the
measured time delay into a robust estimate of the Hubble
constant. Previous attempts in this direction suffered from the
shortcomings of Drizzle and corresponding difficulties in fitting the
PSF \citep{york}. It is our hope that our alternative combining method
can improve these results.

Currently the fitting approach does not include the convolution with
the PSF and the pixel response function. These can be included in a
natural way so that an extended version will be able to `invert' these
effects as well. Additionally, the effect of a gravitational lens also
acts linearly on the brightness distribution, which allows for its
inclusion in the same formalism. With such an overall algorithm, we
will be able to fit the unlensed brightness distribution of the source
directly. The optimal mass distribution of the lens can then be found
by minimizing the remaining residuals. We hope that this approach of a
global fit will be much more reliable and robust than the standard
process of first combining the images, then deconvolving the PSF and
pixel response function with subsequent inversion of the lensing
effect. Similar applications are possible in other fields.

\begin{acknowledgements}
The anonymous referee is thanked for a careful and critical
  review that helped to clarify the presentation of our fitting method
  and its goals. This work was funded by the Emmy-Noether-Programme
of the `Deutsche Forschungsgemeinschaft', reference Wu\,588/1-1.
\end{acknowledgements}

\listofobjects
\bibliographystyle{aa}
\bibliography{paper-aa}
\end{document}